\documentclass[a4paper,oneside,11pt]{article}

\usepackage{jheppub}

\usepackage{amsmath}
\usepackage{amsfonts}
\usepackage{braket}
\usepackage{bbold}
\usepackage{color}
\usepackage{tabularx}

\usepackage{multirow}
\usepackage{hhline}

\usepackage{tikz}
\usetikzlibrary{decorations.pathreplacing}
\usetikzlibrary{decorations.markings}

\usepackage{mathrsfs}
\usepackage{yfonts}

\usepackage{subcaption}
\usepackage{float}
\usepackage{afterpage}
\definecolor{maxcolor}{RGB}{0,166,147}

\DeclareMathOperator{\Tr}{Tr}

\subheader{}
\title{Tripartite information of highly entangled states}

\author{Massimiliano Rota}
\affiliation{ Centre for Particle Theory \& Department of Mathematical Sciences,\\
                     Science Laboratories, South Road, Durham DH1 3LE, UK.}
\emailAdd{massimiliano.rota@durham.ac.uk}

\abstract{Holographic systems require monogamous mutual information for validity of semiclassical geometry. This is encoded by the sign of the tripartite information ($I3$). We investigate the behaviour of $I3$ for all partitionings of systems in states which are highly entangled in a multipartite or bipartite sense. In the case of multipartite entanglement we propose an algorithmic construction that we conjecture can be used to build local maxima of $I3$ for any partitioning. In case of bipartite entanglement we classify the possible values of $I3$ for perfect states and investigate, in some examples, the effect on its sign definiteness due to deformations of the states. Finally we comment on the proposal of using $I3$ as a parameter of scrambling, arguing that in general its average over qubits permutations could be a more sensible measure.}

\begin{document} 

\maketitle
\flushbottom

\section{Introduction}

The \textit{tripartite information} ($I3$) was introduced in \cite{Kitaev:2005dm}, under the name topological entropy, as a quantity to characterize entanglement in states of many-body systems with topological order. Given three subsystems $A$, $B$, $C$ it is defined by the following expression: $I3(A:B:C)=S_A+S_B+S_C-S_{AB}-S_{AC}-S_{BC}+S_{ABC}$, where $S$ is the von Neumann entropy. 

For arbitrary states of many-body systems $I3$ has no definite sign. This is true also in field theory, cf., \cite{Casini:2008wt}. On the contrary, within the context of the gauge gravity duality, it was shown in \cite{Hayden:2011ag} that for states of CFTs with a classical holographic dual, $I3$ is always non-positive. This sign definiteness is a direct consequence of the Ryu-Takayanagi prescription \cite{Ryu:2006bv} for the computation of the von Neumann entropy in holography, and it implies that the holographic mutual information is monogamous.\footnote{More precisely, the proof of monogamy of mutual information refers only to the leading order $N^2$ term of $I3$. In situations where this vanishes (see also \cite{Pakman:2008aa}), order $N^0$ corrections could in principle lead to violation of monogamy \cite{Hayden:2011ag}.}

As consequence of this constraint imposed by holography, the sign of $I3$ has been used in various works to explore what states might be good candidates to encode the properties of classical geometries. In the framework of the ER=EPR\footnote{A conjectured equivalence between entanglement (EPR for Einstein-Podolsky-Rosen) and geometric connectedness (ER for Einstein-Rosen bridges).} proposal \cite{Maldacena:2013xja} for example, it was argued in \cite{Gharibyan:2013aha} that black holes obtained by ``collapsing'' multiple copies of GHZ states of $4$ qubits (for which $I3=+1$) cannot be connected by classical Einstein-Rosen bridges.\footnote{Strictly speaking this was not an holographic argument, as ER=EPR is a general proposal about quantum gravity, nevertheless one can imagine an analogue version of this argument where the geometry is dual to the mentioned qubits state.} The sign of $I3$ was an important consistency check also in the work of \cite{Pastawski:2015qua}, which within the context of the quantum-error-correction interpretation of AdS/CFT of \cite{Almheiri:2014lwa}, built a toy model of holographic states and codes using tensor network constructions. 

For qubits systems, the behaviour of $I3$ was explored in \cite{Rangamani:2015qwa}, where it was shown that random states typically have negative value of $I3$, suggesting that the holographic constraint is not particularly restrictive. Results also indicated that one has to be careful about the particular choices of subsystems for which $I3$ is computed, as some partitionings might be more suitable than others to detect violation of monogamy. Furthermore having holography in mind, the authors proposed that we should look not only at the values of $I3$ for a specific state, but also at how stable this sign is against small deformations or operations performed on it. This proposal was motivated by the finding that for states of 4 qubits, there is only one class of states with definite sign of $I3$.\footnote{States of 4 qubits can be classified into 9 equivalence classes. States within a class are equivalent in the sense that they can be mapped to each other using operations known as SLOCC (stochastic local operations and classical communication). We refer the reader to the original paper for further details.}

Another interesting property of $I3$ which was found in \cite{Rangamani:2015qwa} is the fact that its absolute value seems to be minimized by states which are highly entangled for all bipartitions. In the case of $4$ qubits, a numerical search for the minimum of $I3$ approaches a state, known as M-state in the quantum information literature \cite{Higuchi:2000aa}, which is the maximally entangled state of $4$ qubits. Indeed it was recently shown in \cite{Hosur:2015ylk} that the ``perfect states'' of \cite{Pastawski:2015qua} are the minimizers of $I3$ and that due to this property $I3$ can be used as a measure of information scrambling \cite{Hayden:2007cs}\cite{Sekino:2008he}\cite{Lashkari:2011yi} and quantum chaos \cite{Maldacena:2015waa}.

In this letter we explore the behaviour of $I3$ for some highly entangled states in a bipartite or multipartite sense. In \S\ref{sec:general} we review the definition of $I3$ and discuss some of its general properties. In \S\ref{sec:multipartite} we focus on qubits systems with maximal multipartite entanglement. We explore products of GHZ states and their perturbations in arbitrary directions in Hilbert space, for all possible partitionings of the systems. We move then to the case of states with maximal bipartite entanglement in \S\ref{sec:bipartite}, where we extend the result of \cite{Hosur:2015ylk} to different partitioning of perfect states and comment about their deformations. We conclude in \S\ref{sec:discussion} with a summary and interpretation of the results, together with a discussion about open questions and future directions.

\section{General properties}
\label{sec:general}

\subsection*{Definitions and notation}
\label{}

To simplify the discussion in the following we will focus on generic pure states for systems of an arbitrary number of qu-$b$-its, nevertheless most of the results naturally extend to systems of qu-$d$-its. The fact that we are only looking at pure states will not be a restriction, because for any mixed state one can always consider some purification by enlarging the system. 

Pure states of a system $U$ of $N$ qubits live in a $2^N$ dimensional Hilbert space $\mathcal{H}_{(2^N)}$ with structure $\mathcal{H}_{(2)}^{\otimes N}$, where $\mathcal{H}_{(2)}$ is the two-dimensional Hilbert space of each individual qubit. We will consider subsets of $U$ such that $A\cup B\cup C\subseteq U$ and $A\cap B\cap C=\emptyset$. The Hilbert space corresponding to this partitioning then is $\mathcal{H}_A\otimes\mathcal{H}_B\otimes\mathcal{H}_C$, and the tripartite information is defined as
\begin{align}
I3(A:B:C)\equiv S_A+S_B+S_C-S_{AB}-S_{AC}-S_{BC}+S_{ABC}
\label{}
\end{align}
Since we are only considering pure states of $U$, in the case $A\cup B\cup C=U$ one trivially has $I3\equiv 0$, so in the following we will restrict to $A\cup B\cup C\subset U$. We will use the notation $\mathcal{P}=(A:B:C)$ for a particular partitioning and $I3(\mathcal{P})$ for the tripartite information, stressing that the latter is not only a function of a state but also of a specific partitioning. 

Oftentimes the specific choice of the qubits belonging to the subsets $A$, $B$, $C$ will not be important and we will only need to consider the cardinality of the subsystems. In this case we will write $\mathcal{P}=(a:b:c)$ where $a$, $b$, $c$ refer to the cardinalities of $A$, $B$, $C$ respectively. Ignoring the case $a+b+c=N$ (for which $I3=0$) we then have the conditions
\begin{align}
& 1\leq a\leq N-3, \quad 1\leq b\leq N-3, \quad 1\leq c\leq N-3,\nonumber\\
& 3\leq a+b+c\leq N-1
\label{eq:constraints}
\end{align}
We will use the expression $I3(a:b:c)$ to denote the \textit{set} of all values of $I3(\mathcal{P})$, with $\mathcal{P}=(A:B:C)$, that can be obtained by permuting the specific choice of the qubits in each subset, while keeping $a$, $b$ and $c$ fixed. 

For a given state, or class of states, we want to explore the behaviour of $I3(\mathcal{P})$ for all possible partitionings $\mathcal{P}$.

\subsection*{Equivalences among partitionings}
\label{}

For each partitioning $\mathcal{P}=(A:B:C)$ we will call $D$ the complement of $A\cup B\cup C$ in $U$. As a consequence of the purity of the state of $U$, the entropy of each subsystem is equal to the entropy of the corresponding complementary subsystem. This implies that the tripartite information has the following symmetry \citep{Hosur:2015ylk}
\begin{align}
I3(A:B:C)=I3(A:B:D)=I3(A:C:D)=I3(B:C:D)
\label{eq:symmetry}
\end{align}
As a consequence of Eq.~\eqref{eq:symmetry} then, some of the sets introduced before are actually equivalent. For example, $I3(a:b:c)=I3(N-(a+b+c):b:c)$, see also \citep{Rangamani:2015qwa}. Notice in particular that for the case where $N$ is a multiple of $4$, the set $I3(\frac{N}{4}:\frac{N}{4}:\frac{N}{4})$ is unique. 

\subsection*{Product states}
\label{}

We now explore the behaviour of the tripartite information for states that are obtained by taking products of states of smaller systems. Consider two Hilbert spaces $\mathcal{H}_1, \mathcal{H}_2$ associated to systems $U_1, U_2$ of respectively $N_1$ and $N_2$ qubits. Starting from the states $\ket{\psi}_1\in\mathcal{H}_1$ and $\ket{\phi}_2\in\mathcal{H}_2$ we build the state $\ket{\chi}_{12}=\ket{\psi}_1\otimes \ket{\phi}_2$. We choose then a partitioning $\mathcal{P}_1=(A_1:B_1:C_1)$ of $U_1$ and ask how the values of $I3(\mathcal{P})$ for partitionings of the joint system depend on $I3(\mathcal{P}_1)$ and how the subsets of $U_1$ in $\mathcal{P}_1$ are ``contaminated'' by qubits of $U_2$. This means that we will not change the partitioning of the system $U_1$ but only add qubits of $U_2$ into one or more subsystems of $\mathcal{P}_1$.

Due to the additivity of the entropy for product states, one can check that the following cases are possible
\begin{align}
&\mathcal{P}=(A_1X:B_1:C_1) &\Rightarrow &\quad I3(\mathcal{P})=I3(\mathcal{P}_1)\qquad &&\text{for $X\subseteq U_2$}\nonumber\\
&\mathcal{P}=(A_1X:B_1Y:C_1) &\Rightarrow &\quad I3(\mathcal{P})=I3(\mathcal{P}_1)\qquad &&\text{for $X\cup Y\subseteq U_2$}\nonumber\\
&\mathcal{P}=(A_1X:B_1Y:C_1Z) &\Rightarrow &\quad I3(\mathcal{P})=I3(\mathcal{P}_1)+I3(\mathcal{P}_2)\qquad &&\text{for $X\cup Y\cup Z\subset U_2$}\nonumber\\
&\mathcal{P}=(A_1X:B_1Y:C_1Z) &\Rightarrow &\quad I3(\mathcal{P})=I3(\mathcal{P}_1)\qquad &&\text{for $X\cup Y\cup Z=U_2$}
\label{eq:product}
\end{align}
where $\mathcal{P}_2=(X:Y:Z)$. In this set-up then, $I3(\mathcal{P})$ is either invariant or additive. We will come back to this property and some of its consequences in the following sections.

\subsection*{General bounds}
\label{}

We first look at general bounds for $I3(\mathcal{P})$ that are satisfied by all states and partitionings. In the next sections we will explore further bounds that apply to specific partitionings for different classes of states. The fact that $I3(\mathcal{P})$ is in general bounded is an obvious consequence of the bound of the entropy. 

A lower bound for the tripartite information was given in \citep{Hosur:2015ylk} and can be found by rewriting $I3(\mathcal{P})$ as\footnote{We thank Beni Yoshida for a clarification about this point.} 
\begin{align}
I3(A:B:C)=I(A:B)+I(A:C)-I(A:BC)
\label{eq:I3_with_mutual_info}
\end{align}
where $I(X:Y)=S_X+S_Y-S_{XY}$ is the mutual information. From the non-negativity of mutual information it follows then that $I3(A:B:C)\geq -I(A:BC)$. Furthermore $I(A:BC)\leq 2\min(S_A, S_{BC})$ which implies $I3(A:B:C)\geq -2\min(S_A, S_{BC})$. One can then repeat the same argument using the symmetry Eq.~\eqref{eq:symmetry}, getting
\begin{align}
I3(A:B:C)\geq -2\min(S_A, S_B, S_C, S_D, S_{AB}, S_{AC}, S_{AD}, S_{BC}, S_{BD}, S_{CD})
\end{align}
Note that the minimal value of $I3(\mathcal{P})$ is attained for states such that $S_{XY}\geq S_X\; \forall X,Y$. In this case the bound is the one reported in \citep{Hosur:2015ylk}.
\begin{align}
I3(A:B:C)\geq -2\min(S_A, S_B, S_C, S_D)
\label{eq:lower_bound}
\end{align}
When $N$ is a multiple of $4$, $I3(\mathcal{P})$ is minimized by states such that all the entropies $S_X$ are maximal and $\mathcal{P}=(\frac{N}{4}:\frac{N}{4}:\frac{N}{4})$; in this case $I3(\mathcal{P})=-\frac{N}{2}$. We will analyse the behaviour of $I3(\mathcal{P})$ for these states in more detail in \S\ref{sec:bipartite}. For $N=1,2,3\; (\text{mod}\,4)$ instead, the bound would be tighter. 

To derive an upper bound one could start again from Eq.~\eqref{eq:I3_with_mutual_info}, but using strong subadditivity (SSA)\footnote{For the convenience of the reader we report here the definition of strong subadditivity $S_A+S_B\leq S_{AC}+S_{BC}$.} the bound is more restrictive. We can simply rewrite the tripartite information as
\begin{align}
I3(A:B:C)\equiv & \frac{1}{2}\left(S_A+S_B-S_{AC}-S_{BC}\right)+\frac{1}{2}\left(S_A+S_C-S_{AB}-S_{CB}\right)\nonumber\\
+ & \frac{1}{2}\left(S_B+S_C-S_{BA}-S_{CA}\right)+S_{ABC}\equiv \Sigma_{ABC}+S_{ABC}
\label{}
\end{align}
SSA implies then $\Sigma_{ABC}\leq 0$. Using purity of the global state (which implies $S_{ABC}=S_D$) and the symmetry Eq.~\eqref{eq:symmetry} one gets
\begin{align}
I3(A:B:C)\leq \min(S_A,S_B,S_C,S_D)
\label{eq:upper_bound}
\end{align}
Similarly to before, when $N$ is a multiple of $4$, $I3(\frac{N}{4}:\frac{N}{4}:\frac{N}{4})$ is maximal for states with maximal entropies $S_X$. In this case $I3(\mathcal{P})\leq \frac{N}{4}$.

\section{States with maximal multipartite entanglement}
\label{sec:multipartite}

The GHZ state of $N$ qubits is defined as
\begin{align}
\ket{\text{GHZ}_N}=\frac{1}{\sqrt{2}}\left(\ket{0...0}+\ket{1...1}\right)
\end{align}
and it is a well known example of a state for which $I3(\mathcal{P})\geq 0$. Ignoring the trivial case $N=3$ for which $I3(\mathcal{P})=0$, an immediate calculation shows that for any subsystem $X$ of the $N$ qubits, the entropy is $S_X=1$. This implies that for any partitioning $\mathcal{P}$, one has $I3(\mathcal{P})=1$ for any $N$. For the case $N=4$ this immediately implies that the state GHZ$_{4}$ is the global maximum of $I3(\mathcal{P})$, because it saturates the bound Eq.~\eqref{eq:upper_bound}. 

Consider now the state $\ket{\text{GHZ}_4}^{\otimes k}$, obtained by taking a tensor product of $k$ copies of the state GHZ$_{4}$. For this state of the new $N=4k$ qubits system we look at the partitioning defined as follows: take one qubit for each copy of the GHZ$_{4}$ state and put it into the subsystem $A$ of the larger system, then repeat the same procedure for subsystems $B$ and $C$. For this particular partitioning it follows from Eq.~\eqref{eq:product} that $I3(\mathcal{P})=k=\frac{N}{4}$. As before, this value saturates the bound Eq.~\eqref{eq:upper_bound}, implying that these product states are the global maxima of $I3(\frac{N}{4}:\frac{N}{4}:\frac{N}{4})$ for $4k$ qubits.

In this section we discuss how the values of $I3(\mathcal{P})$ depend on the different partitionings $\mathcal{P}$ for deformations of GHZ$_N$ states. In particular we present an algorithmic construction that we conjecture can be used to build local maxima of $I3(\mathcal{P})$ for arbitrary $N$ and any given $\mathcal{P}$. In the particular case $N=4k$ this construction recovers the previous result for the state $\ket{\text{GHZ}_4}^{\otimes k}$ and generates an entire new family of states that saturate the bound.

\subsection*{Deformations of GHZ$_N$ states}

We start by considering the following deformation of the GHZ$_N$ state
\begin{align}
\ket{\text{GHZ}_N}\rightarrow \ket{\psi^\epsilon_I}=
\begin{cases}
\frac{1}{\sqrt{1+|1+\epsilon|^2}}\left(\ket{0...0}+\ket{1...1}+\epsilon\ket{I}\right) & \text{if $\ket{I}\in\{\ket{0...0},\ket{1...1}\}$}\\
\frac{1}{\sqrt{2+|\epsilon|^2}}\left(\ket{0...0}+\ket{1...1}+\epsilon\ket{I}\right) & \text{otherwise}\\
\end{cases}
\end{align}
where $\ket{I}$ is an element of the computational basis $\{\ket{0...0}, \ket{0...1}, ... \ket{1...1}\}$. Consider then a generic bipartition of the system into a subsystem $X$ of size $x$ and its complement $X^c$ of size $N-x$. The reduced density matrix $\rho_X$ associated to the subsystem $X$ is given by (up to the normalization factor)
\begin{align}
\rho_X(\epsilon,I)\equiv \Tr_{X^c}{\rho^\epsilon_I}= & \ket{0...0}\bra{0...0}+\ket{1...1}\bra{1...1}+|\epsilon|^2\ket{I_X}\bra{I_X}\nonumber\\
&+\begin{cases}
\epsilon^*\ket{0...0}\bra{I_X}+\epsilon\ket{I_X}\bra{0...0} & \text{if $\ket{I_{X^c}}$ is Homogeneous in 0's}\\
\epsilon^*\ket{1...1}\bra{I_X}+\epsilon\ket{I_X}\bra{1...1} & \text{if $\ket{I_{X^c}}$ is Homogeneous in 1's}\\
0 & \text{if $\ket{I_{X^c}}$ is not Homogeneous}\\
\end{cases}
\label{eq:rho_i1}
\end{align}
where $\ket{I_X}$ and $\ket{I_{X^c}}$ are the states of subsystems $X$ and $X^c$ when the global system is in the state $\ket{I}$. By the expression ``Homogeneous in $0$'s'' we mean $\ket{I_{X^c}}=\ket{0}^{\otimes N-x}$ (and similarly for $1$'s). $\ket{I_{X^c}}$ instead is not Homogeneous if $\ket{I_{X^c}}=\ket{0}^{\otimes \gamma}\otimes\ket{1}^{\otimes\delta}$ for any $\gamma,\delta$ such that $\gamma+\delta=N-x$. In the following we will use short expressions like ``$I_{X^c}$ is $Hom$'' to indicate these cases (eventually dropping also the ``ket'', as we think about $I_{X^c}$ simply as a string of digits).

Depending on the homogeneity properties of $\ket{I_X}$ we then have four possibilities for the final expression of the reduced density matrix. We list the possible cases, together with the corresponding eigenvalues of $\rho_X(\epsilon,I)$, in Tab.~\ref{tab:ghz_deformations_entropies}. This is an exact result, not only perturbative in $\epsilon$.

\begin{table}[tb]
\centering
\begin{tabular}{|l|l|l|}
\hline
$S_1(\epsilon)$ & both $I_X$ and $I_{X^c}$ are $Hom$ in $\eta$ & $\lambda_1=\frac{(1+2\text{Re}\epsilon+|\epsilon|^2)}{1+|1+\epsilon|^2},\;\lambda_2=\frac{1}{1+|1+\epsilon|^2}$\\
\hline
$S_2(\epsilon)$ & $I_X$ is $Hom$ in $\eta$ and $I_{X^c}$ in $\bar{\eta}$ & $\lambda_{12}=\frac{(2+|\epsilon|^2\pm|\epsilon|\sqrt{4+|\epsilon|^2})}{2(2+|\epsilon|^2)}$\\
\hline
$S_3(\epsilon)$ & either $I_X$ or $I_{X^c}$ is $Hom$ & $\lambda_1=\frac{1}{2+|\epsilon|^2},\;\lambda_2=\frac{1+|\epsilon|^2}{2+|\epsilon|^2}$\\
\hline
$S_4(\epsilon)$ & both $I_X$ and $I_{X^c}$ are not $Hom$ & $\lambda_1=\frac{1}{2+|\epsilon|^2},\;\lambda_2=\frac{1}{2+|\epsilon|^2},\;\lambda_3=\frac{|\epsilon|^2}{2+|\epsilon|^2}$\\
\hline
\end{tabular}
\caption{The table shows the four possible configurations of the strings of digits $I_X$ and $I_{X^c}$ and the set of eigenvalues of the corresponding expression for the reduced density matrix $\rho_X$. The functions $S_i(\epsilon)$ are the entropies, for the various cases labelled by $i$. The parameters $\eta,\bar{\eta}$ are mutually exclusive variables, when $\eta=0$, $\bar{\eta}=1$, and vice versa.}
\label{tab:ghz_deformations_entropies}
\end{table}

The functions $S_i(\epsilon)$ that give the entropy of $\rho_X(\epsilon,I)$ depending on its possible structures, all have vanishing first derivative at $\epsilon=0$. This shows that in the Hilbert space of $N$ qubits, and for any $N$, the state GHZ$_N$ is a saddle point of $I3(\mathcal{P})$ for all $\mathcal{P}$.\footnote{This result immediately follows from the fact that for any $\mathcal{P}$ the tripartite information is just a linear combination of entropies.} Furthermore, the functions $S_1(\epsilon)$, $S_2(\epsilon)$ and $S_3(\epsilon)$ are all decreasing, while $S_4(\epsilon)$ is increasing. In particular $S_3(\epsilon)$ decreases only at order $\epsilon^4$.

With the set of possible entropies at hand, we now want to classify the possible behaviours of the tripartite information of $\ket{\psi^\epsilon_I}$, depending on the partitioning and the direction of the deformation $\ket{I}$. A natural classification would proceed by first fixing a partitioning $\mathcal{P}$, and then looking at the behaviour of $I3(\mathcal{P})$ in all possible directions $\ket{I}$. Nevertheless, due to the nature of the problem, it is more natural to proceed in the opposite way. We first fix a direction $\ket{I}$ of deformation and then derive the behaviour of $I3(\mathcal{P})$ for all possible $\mathcal{P}$. This is more natural because the behaviour of $I3(\mathcal{P})$ will just depend on the homogeneity properties of the strings $I_A, I_B, I_C, I_D$ derived from $\ket{I}$ under $\mathcal{P}$, and the analogous properties for their unions.\footnote{Recall that for two $Hom$ strings $X,Y$ the union is not $Hom$ if $X$ is $Hom$ in $1$'s (or $0$'s) and $Y$ in $0$'s ($1$'s).} The possible cases are shown in Tab.~\ref{tab:ghz_deformations_I3} and are classified using a parameter $\phi$ that counts the number of strings $X\in\{I_A,I_B,I_C,I_D\}$ which are $Hom$.

\begin{table}[tb]
\centering
\begin{tabular}{|c|c|c|}
\hline
$\phi$ & Details of $I_A,I_B,I_C,I_D$ & $I3(\mathcal{P})$ \\
\hline
0 & $X$ is not $Hom,\;\forall X$ & $S_4(\epsilon)$\\
\hline
1 & $\exists!\,X$ that is $Hom$ & $S_3(\epsilon)$\\
\hline
\multirow{2}{*}{2} & $X,Y$ are $Hom$ in $\lambda$ & $S_3(\epsilon)$\\
\hhline{~--}
 & $X$ is $Hom$ in $\eta$ and $Y$ is $Hom$ in $\bar{\eta}$ & $2S_3(\epsilon)-S_4(\epsilon)$\\
\hline
 \multirow{2}{*}{3} & $X$ is not $Hom$ and $X^c$ is $Hom$ & $S_3(\epsilon)$\\
 \hhline{~--}
 & $X$ is not $Hom$ and $X^c$ is not $Hom$ & $2S_3(\epsilon)-S_4(\epsilon)$\\
 \hline
\multirow{3}{*}{4} & $I_A\cup I_B\cup I_C\cup I_D\equiv I$ is $Hom$ & $S_1(\epsilon)$\\
 \hhline{~--}
 & $X$ is $Hom$ in $\eta$ and $X^c$ is $Hom$ in $\bar{\eta}$ & $S_2(\epsilon)$\\
 \hhline{~--}
 & $X\cup Y$ is $Hom$ in $\eta$ and $(X\cup Y)^c$ is $Hom$ in $\bar{\eta}$ &  $4S_3(\epsilon)-2S_4(\epsilon)-S_2(\epsilon)$ \\
 \hline
\end{tabular}
\caption{The table lists the possible behaviour of $I3(\mathcal{P})$ for different $\mathcal{P}$ and a fixed direction of deformation $\ket{I}$. The parameter $\phi$ is the number of strings among $I_A,I_B,I_C,I_D$ which are $Hom$ in $1$'s or $0$'s. As in Tab.~\ref{tab:ghz_deformations_entropies}, $\eta$ and $\bar{\eta}$ are mutually exclusive variables, when $\eta=0$, $\bar{\eta}=1$, and vice versa.}
\label{tab:ghz_deformations_I3}
\end{table}

The results of Tab.~\ref{tab:ghz_deformations_I3} show that for a given direction $\ket{I}$, $I3(\mathcal{P})$ of GHZ$_N$ can increase only for those $\mathcal{P}$ such that all the strings $I_A,I_B,I_C,I_D$ are not $Hom$. Since a string made of a single digit is always $Hom$, the following lemma follows

\paragraph{Lemma:} \textit{For any $N$, the GHZ$_N$ state is a local maximum of $I3(\mathcal{P})$ for any $\mathcal{P}$ such that at least one of the subsystems contains only a single qubit.}

Since for $N\leq 7$  this always happens, in this case the GHZ$_N$ state is a local maximum of $I3(\mathcal{P})$ for all $\mathcal{P}$.

For arbitrary $N$ and $\mathcal{P}$ instead, the GHZ$_N$ states are not local maxima. Nevertheless, since we know exactly how the value of $I3(\mathcal{P})$ behaves along each direction (not only perturbatively), for fixed $\mathcal{P}$ we can choose a direction $\ket{I_1}$ along which $I3(\mathcal{P})$ grows and follow it until we reach a maximum in that direction. One can check that the function $S_4(\epsilon)$ reaches a maximum along $\ket{I_1}$ for $|\epsilon|=1$. We can then build the new state 
\begin{align}
\ket{\text{GHZ}_N}\rightarrow \ket{\psi_1}=\frac{1}{\sqrt{3}}\left(\ket{0...0}+\ket{1...1}+e^{i\theta_1}\ket{I_1}\right)
\label{eq:psi1}
\end{align}
This new state of course is not guaranteed to be a local maximum of $I3(\mathcal{P})$. To investigate whether this is the case or not, we can again look at deformations along all possible directions. We then build the new state
\begin{align}
\ket{\psi_1}\rightarrow \ket{\psi_2^\epsilon}=\frac{1}{\sqrt{\mathcal{N}}}\left(\ket{0...0}+\ket{1...1}+e^{i\theta_1}\ket{I_1}+\epsilon\ket{I_2}\right)
\label{eq:psi2}
\end{align}
For an arbitrary bipartition of the system into $X$ and $X^c$, the reduced density matrix $\rho_X(\epsilon,I_1,I_2)$ will have the following structure (up to normalization factors)
\begin{align}
\rho_X(\epsilon,I_1,I_2)=\rho_X(e^{i\theta_1},I_1)+\rho_X(\epsilon,I_2)+e^{i\theta_1}\epsilon^*\ket{I_X^1}\bra{I_X^2}+e^{-i\theta_1}\epsilon\ket{I_X^2}\bra{I_X^1}
\label{eq:rho_i1_i2}
\end{align}
In Eq.~\eqref{eq:rho_i1_i2} the expressions $\rho_X(e^{i\theta_1},I_1)$ and $\rho_X(\epsilon,I_2)$ correspond to matrices of the form Eq.~\eqref{eq:rho_i1}, with deformations along $\ket{I_1},\ket{I_2}$ and coefficients respectively $e^{i\theta_1}$ and $\epsilon$. The last two terms are ``interference'' terms that survive only when $\ket{I_{X^c}^1},\ket{I_{X^c}^2}$ (defined as in Eq.~\eqref{eq:rho_i1}) are not orthogonal.

We check numerically for many examples that the interference terms reduce the entropy, while the entropy increases if these terms disappear. This observation motivates the following construction. Given a partitioning $\mathcal{P}=(A:B:C:D)$ for a system of $N$ qubits, start with the GHZ$_N$ state. Then pick a direction $\ket{I_1}$ with the property that all the strings $I_A^1,I_B^1,I_C^1,I_D^1$ are not $Hom$, such that $I3(\mathcal{P})$ will grow, and build the new state Eq.~\eqref{eq:psi1}. Then look for a second possible direction $\ket{I_2}$ such that $I_A^2,I_B^2,I_C^2,I_D^2$ are again not $Hom$ and $\braket{I_A^1|I_A^2}=\braket{I_B^1|I_B^2}=\braket{I_C^1|I_C^2}=\braket{I_D^1|I_D^2}=0$, and build the new state Eq.~\eqref{eq:psi2} with $\epsilon=e^{i\theta_2}$. Finally iterate this construction for all possible directions that satisfy these conditions. This procedure is limited by the subset $X\in\{A,B,C,D\}$ which has minimal size $x$, and will stop at some point. We then conjecture the following

\paragraph{Conjecture:}\textit{All the states that can be built following this algorithmic construction are local maxima of $I3(A:B:C:D)$.} 

On can check for example that in the case $N=4k$, for specific permutation of the qubits in the partitioning $\mathcal{P}=(\frac{N}{4}:\frac{N}{4}:\frac{N}{4}:\frac{N}{4})$, and picking all the phases to be $e^{i\theta_i}=1$, the procedure starts with the state $\ket{\text{GHZ}_N}$ and ends with the state $\ket{\text{GHZ}}_4^{\otimes k}$, recovering the result stated before. We leave the general proof of this conjecture as an open problem for future work.

\section{States with maximal bipartite entanglement}
\label{sec:bipartite}

In this section we focus on bipartite entanglement and investigate the behaviour of the tripartite information for states that are highly entangled for all possible bipartitions of the system. The search for this kind of states, usually called MMES (maximal multi-qubit entangled states),\footnote{They are sometimes called \textit{maximal multipartite entangled states}, but this denomination might be misleading, suggesting some connection to multipartite entanglement. Instead, ``multipartite'' here refers to the fact that we are looking not only at entanglement for one particular bipartition of the system, but for all bipartitions.} is an important problem in quantum information theory \cite{Facchi:2007aa}, where entanglement is a resource for the implementation of many protocols.

A particularly interesting subclass of MMES are the \textit{perfect} MMES, for which the entropy of each subsystem is exactly maximal; these are indeed the \textit{perfect states} of \citep{Pastawski:2015qua} and \citep{Hosur:2015ylk}. In the case of qubits it is known that they do not exist for $N\geq 8$ \citep{Scott:2003aa}. For qudits, examples can be found using stabilizer code \cite{Gottesman:1997aa} techniques \citep{Hosur:2015ylk}\cite{Yang:2015uoa}.

We want to explore the behaviour of $I3(\mathcal{P})$ for different partitionings of these states. We start with perfect states, for which a classification of the possible values of $I3(\mathcal{P})$ is possible even without knowing an explicit expression. Next we investigate some examples of MMES for $N=2,4,6,8$ and some other states that can be built from them.

\subsection*{Perfect states}

\begin{table}[tb]
\centering
\begin{tabular}{|c|c|c|c|c|c|c|}
\hline
$\exists X,\;|X|\geq\frac{N}{2}$ & \multicolumn{6}{c|}{$I3(\mathcal{P})=0,\;\forall\mathcal{P}$} \\
\hhline{|=|=|=|=|=|=|=|}
\multirow{5}{*}{$|X|<\frac{N}{2},\;\forall X$} & $\chi$ & $I3$ & $\mathcal{P}_{\min}$ & $I3_{\min}$ & $\mathcal{P}_{\max}$ & $I3_{\max}$\\
\hhline{~------}
 & $0$ & $-2\alpha$ & $a=b=c=\frac{N}{4}$ & $-\frac{N}{2}$ & $\alpha=1$ & $-2$ \\
 \hhline{~------}
 & $1$ & $-2c$ & $a=b-1=c+1=\frac{N}{4}$ & $-\frac{N}{2}+2$ & $c=1$ & $-2$ \\
\hhline{~------}
 & $2$ & $-N+2a$ & $a-1=b=c=\frac{N}{4}$ & $-\frac{N}{2}+2$ & $a=\frac{N}{2}-1$ & $-2$ \\
\hhline{~------}
 & $3$ & $2\alpha-N$ & $a-1=b-1=c=\frac{N}{4}$ & $-\frac{N}{2}+4$ & $\alpha=\frac{N}{2}-1$ & $-2$ \\
\hline
\end{tabular}
\caption{The table shows the classification of the values of $I3(\mathcal{P})$ for perfect states, for all possible partitionings of the system. When a subsystem $X$ (possibly also $X=D$) contains at least half of the qubits, $I3(\mathcal{P})$ vanishes. The other cases are classified according to the parameter $\chi$ defined in Eq.~\eqref{eq:parameter_chi}. For each case the value of $I3(\mathcal{P})$ is given as a function of $(a,b,c)$. Maximal and minimal values of $I3(\mathcal{P})$ and the corresponding partitionings are also shown for each case. The parameter $\alpha$ is defined as $\alpha=a+b+c-\frac{N}{2}$.}. 
\label{tab:perfect_states}
\end{table}

Perfect states are defined as those states for which each subsystem $X\subseteq U$ (with $|X|=x$) has exactly maximal entropy
\begin{align}
S_x= \begin{cases}
x &\text{for $x\leq\frac{N}{2}$} \\
N-x &\text{for $x>\frac{N}{2}$}\\
\end{cases}
\label{eq:perfect_entropy}
\end{align}
Since perfect states are symmetric under permutations of the qubits, we can classify the behaviour of $I3(\mathcal{P})$ looking at the sets $I3(a:b:c)$ with constraints Eq.~\eqref{eq:constraints} on $a$, $b$ and $c$. Once the sizes of subsystems are specified, the entropies are given by Eq.~\eqref{eq:perfect_entropy} and we can immediately compute the value of $I3(\mathcal{P})$. For simplicity, in the following we will assume that $N$ is a multiple of $4$.

When $a+b+c<\frac{N}{2}$, or when any of the subsystems contains $\frac{N}{2}$ qubits or more, one has $I3(\mathcal{P})=0$. The two cases are equivalent because of Eq.~\eqref{eq:symmetry}, indeed when $a+b+c<\frac{N}{2}$, it follows that $d\geq\frac{N}{2}$. To classify all other possible cases we will use a parameter $\chi$, defined as the number of unions of two subsystems $X,Y$ that contain at least $\frac{N}{2}$ qubits, i.e.  $|X\cup Y|\geq\frac{N}{2}$. To simplify the notation, and without loss of generality, we assume that $a\geq b\geq c$, such that 
\begin{align}
\chi= 
\begin{cases}
0 &\text{for $|X\cup Y|<\frac{N}{2},\;\forall X,Y$} \\
1 &\text{for $|A\cup B|\geq\frac{N}{2}$ but $|A\cup C|,|B\cup C|<\frac{N}{2}$}\\
2 &\text{for $|A\cup B|,|A\cup C|\geq\frac{N}{2}$ but $|B\cup C|<\frac{N}{2}$}\\
3 &\text{for $|X\cup Y|\geq\frac{N}{2},\;\forall X,Y$} \\
\end{cases}
\label{eq:parameter_chi}
\end{align}
The classification of the possible values of $I3(\mathcal{P})$ is summarized in Tab.~\ref{tab:perfect_states}, where we also indicate the specific partionings that maximize or minimize the value of $I3(\mathcal{P})$ in each case. 

Note that the partitioning $\mathcal{P}=(\frac{N}{4}:\frac{N}{4}:\frac{N}{4})$ is the minimizer of $I3(\mathcal{P})$ for perfect states. Furthermore, since in this case $I3(\mathcal{P})=-\frac{N}{2}$, perfect states saturate the bound Eq.~\eqref{eq:lower_bound} and are absolute minima of $I3(\mathcal{P})$. Indeed, this motivated the proposal of \citep{Hosur:2015ylk} that $I3(\mathcal{P})$ can be used as a parameter for scrambling.

Suppose now that for some value of $N$ (again multiple of $4$), a perfect state $\ket{\text{P}_N}$ exists. Then we can take two copies of this state and build a new state of a system of size $2N$ taking the product $\ket{\text{P}_N}\otimes \ket{\text{P}_N}$. This new state would not be a perfect state any more, nevertheless according to the additivity of $I3(\mathcal{P})$ shown in Eq.~\eqref{eq:product}, there is some partitioning that gives $I3(\mathcal{P}_{2N})=2\times I3(\mathcal{P}_N)=-\frac{(2N)}{2}$. This simple fact shows that although it is true that a scrambled state would minimize $I3(\mathcal{P})$ of a partitioning $\mathcal{P}=(\frac{N}{4}:\frac{N}{4}:\frac{N}{4})$, the converse is not true. Only if we know that the state we are dealing with is completely symmetric under all permutations, the value of $I3(\mathcal{P})$ is sufficient to imply scrambling. 

Finally, we comment on another interesting property that emerges from the results of Tab.~\ref{tab:perfect_states}. Note that while the lower bound of $I3(\mathcal{P})$ for different partitionings scales with $N$, the upper bound does not. In particular there are partitionings for which $I3(\mathcal{P})=0$. In the holographic perspective, these are the ones we should be more careful about, as they get closer to the violation of monogamy for mutual information. It would be interesting to study the behaviour of perfect states for such partitionings under the effect of arbitrary operations performed on the constituents of the system. We leave the general question for future work, while in the next section we explore the example of $N=6$, for which a perfect state of qubits exists and is known explicitly.

\subsection*{Some examples of MMES states}

We now explore the behaviour of the tripartite information for systems of $N=2,4,6,8$ qubits, focusing on highly entangled states and some deformations of them. We also compare the value of $I3(\mathcal{P})$ to the value obtained for particular product states, suggesting that the average $\overline{I3(\mathcal{P})}$ over permutation of the qubits could be a more sensible measure to evaluate scrambling.

\paragraph*{N=2} Obviously $I3(\mathcal{P})$ for states of just $2$ qubits is nonsense. Starting with maximally entangled states $\ket{M_2}$ of $2$ qubits (Bell pairs), we can build maximally entangled states of an arbitrary even number of qubits by simply taking the product $\ket{M_2}^{\otimes k}$. These states are indeed maximally entangled but only for certain bipartitions. In particular there is only one subsystem containing $\frac{N}{2}$ qubits which has maximal entropy. For the case $k=2$ one gets a maximally entangled state of $4$ qubits for which $I3(\mathcal{P})=0$. As a consequence of Eq.~\eqref{eq:product} when we take a product with a new copy of $\ket{M_2}$, $I3(\mathcal{P})$ is invariant. By induction one has $I3(\mathcal{P})=0$ for arbitrary $k$. In other words, any ``distilled'' state\footnote{Distillation is the process of extraction of Bell pairs from a given state using LOCC operations.} has $I3(\mathcal{P})=0$ for all $\mathcal{P}$. The converse is obviously not true, a product state for all qubits contains no entanglement and would equally have $I3\equiv 0$.

\paragraph*{N=4} The MMES of $4$ qubits  was found in \cite{Higuchi:2000aa} and is known as M state. It has the form 
\begin{align}
\ket{M_4}=\ket{0011}+e^{-\frac{\pi}{3}i}\ket{0101}-e^{\frac{\pi}{3}i}\ket{0110}-e^{\frac{\pi}{3}i}\ket{1001}+e^{-\frac{\pi}{3}i}\ket{1010}+\ket{1100}
\end{align}
Although this is the maximally entangled state of 4 qubits, it is not a perfect state as the entropies of one and two qubits are respectively $S_{\{1\}}=1$, $S_{\{2\}}=\frac{1}{2}\log_2{12}\approx 1.79248<2$. The tripartite information for this state has value $I3(\mathcal{P})=4-\frac{3}{2}\log_2{12}\approx -1.37744$. By deforming the state with a small pertubation in any direction in Hilbert space, one can check numerically that this state is a local minimum for $I3(\mathcal{P})$.

\paragraph*{N=6} In the particular case of 6 qubits the perfect state $\ket{\text{P}_6}$ is known explicitly,\footnote{We refer the reader to the original paper for its expression.} it was found in \cite{Borras:2008aa}. We can then investigate the effect of deformations of the state on the sign of $I3(\mathcal{P})$. Following the classification of Tab.~\ref{tab:perfect_states}, we can look for the partitionings for which $I3(\mathcal{P})=0$. We have the possible cases $\mathcal{P}=(1:1:1)$ or $\mathcal{P}=(3:1:1)$, but they are equivalent according to Eq.~\eqref{eq:symmetry}. Starting with the state $\ket{\text{P}_6}$ we can deform it in the directions labelled by the computational basis: $\ket{\psi^\epsilon_I}=\ket{\text{P}_6}+\epsilon\ket{I}$. A numerical check shows that $I3(1:1:1)$ decreases in all directions; small perturbations cannot change its sign. We can also explore the effect of measurements performed on some of the qubits of the system. We can for example measure a single qubit with any of $\sigma_x,\sigma_y,\sigma_z$ or we can do a Bell measurement and project two qubits onto a maximally entangled state. In both these cases one can check that for the states obtained under these operations it is still true that $I3(\mathcal{P})\leq 0$ for all $\mathcal{P}$.

\paragraph*{N=8} An $8$ qubits MMES was found in \cite{Zha:2012aa}, we will refer to it as the $\ket{\text{M}_8}$ state. As for $N=4$, a numerical check shows that this state is a local minimum of $I3(2:2:2)$ in Hilbert space. In particular $I3(2:2:2)[\text{M}_8]\approx-1.35458$, while for a perfect state of $8$ qubits ($\ket{\text{P}_8}$ which does not exist) it would have been $I3(2:2:2)[\text{P}_8]=-4$. We can now compare this result with the value of $I3(2:2:2)$ for the state $\ket{\text{M}_4}\otimes\ket{\text{M}_4}$, where $\ket{\text{M}_4}$ is the MMES of $4$ qubits introduced before. In this case one has $I3(2:2:2)[\text{M}_4\otimes\text{M}_4]\approx -2.75489<-1.35458$. This simple observation suggests again\footnote{See also the discussion about perfect states.} that one should be careful in using $I3(\mathcal{P})$ as a parameter of scrambling. On the other hand, since this value of $I3(2:2:2)[\text{M}_4\otimes\text{M}_4]$ is only attained for some permutations of the qubits, one can ask whether the average value $\overline{I3(2:2:2)}$ over all permutation is a more sensible measure. The state $\ket{\text{M}_8}$ is completely symmetric under permutations of the qubits, so that the average tripartite information has the same value obtained before. This is not true for the state $\ket{\text{M}_4}\otimes\ket{\text{M}_4}$ in which case, taking into account the combinatorics,\footnote{For the state $\ket{\text{M}_4}\otimes\ket{\text{M}_4}$, the tripartite information is either $-2.75489$ or $0$. There are in general $420$ possible qubits permutations corresponding to the partitioning $\mathcal{P}=(2:2:2)$ of the system, $96$ of which give the non vanishing value.} one gets $\overline{I3(2:2:2)}[\text{M}_4\otimes\text{M}_4]\approx -0.62969>-1.35458$. For $N=8$ a perfect state does not exist and it is natural to consider the MMES as the scrambled state in this Hilbert space. This example then shows that the MMES is not the absolute minimizer for a single value of $I3(\mathcal{P})$ corresponding to a specific permutation of the qubits . On the other hand the average $\overline{I3(\mathcal{P})}$ seems to be minimized by the MMES.

\section{Discussion}
\label{sec:discussion}

In this letter we explored the behaviour of the tripartite information for different partitionings of systems in highly entangled states. For simplicity we focused in particular on systems of qubits, but most of the result can be generalized to constituents that live in a higher dimensional Hilbert space, i.e. qudits.

After a discussion about general properties of $I3(\mathcal{P})$, we started by looking at states that maximize multipartite entanglement, namely GHZ$_N$ states. We showed how $I3(\mathcal{P})$ changes for deformations of the states in various directions in Hilbert space, depending on the different partitionings of the system. Then we proposed an algorithmic construction that we conjectured can be used to build local maxima of $I3(\mathcal{P})$ for arbitrary $N$ and $\mathcal{P}$. We leave the proof of this conjecture and the extension to higher dimensional generalizations of GHZ$_N$ states for future work.

Next we moved to states that manifest a high amount of bipartite entanglement for all possible bipartitions of the system. We explored the general behaviour of the perfect states of \citep{Pastawski:2015qua} for all possible partitionings and then looked at some examples of qubits states which although not perfect, are known to be highly entangled for all bipartitions. 

Our main motivation for studying the tripartite information came from holography, where $I3(\mathcal{P})$ has definite non-positive sign and captures the monogamy of mutual information \citep{Hayden:2011ag}. Drawing from the results of the previous sections, we conclude with some observations which are relevant in the holographic context, posing some open questions that we leave to future investigations. 

\paragraph{The sign of the tripartite information} The work of \citep{Rangamani:2015qwa} asked the question of how generic is monogamy of mutual information, and consequently how restrictive is the constraint imposed by holography. It was found numerically that for random states of $6$ and $8$ qubits it is extremely difficult to obtain states with positive value of $I3(\mathcal{P})$. Furthermore, it was observed that when $\mathcal{P}=(1:1:1)$, the values of $I3(\mathcal{P})$ for random states, although still negative, approach $I3(\mathcal{P})=0$. This matches with the behaviour of perfect states shown in Tab.~\ref{tab:perfect_states}, which under the same assumptions for $\mathcal{P}$, have precisely $I3(\mathcal{P})=0$. 

This similarity between the distribution of random states for different choices of $\mathcal{P}$ and the values of $I3(\mathcal{P})$ for perfect states, extends to all cases where the size of subsystems in $\mathcal{P}$ is much smaller (or much larger) than half of the size of the entire system. This can be interpreted as a consequence of Page theorem \cite{Page:1993df}, which precisely under the same assumptions for the size of subsystems, implies that random states are almost maximally entangled. It would be interesting to explore further the relation between random and perfect states. In particular, since as far as entropies are concerned, they generically have a similar behaviour, one could try to make this connection quantitative by introducing a notion of ``typicality''\footnote{\textit{Typicality} here has to be interpreted in the sense of \cite{Lebowitz:1999aa}. According to some measure, the distance between the behaviour of random and perfect states would be exponentially suppressed for large $N$.} for perfect states.

Next, since for certain partitionings of perfect states one gets $I3(\mathcal{P})=0$, it is natural to ask how stable is the sign definiteness of $I3(\mathcal{P})$ for these particular partitionings when we deform the states either by some perturbation or by some operation performed on the constituents. Without a general expression at hand for perfect states, we focused on the example of the $6$ qubits systems, for which the perfect state is known explicitly. We checked numerically that any deformation in any direction in Hilbert space can only decrease the value of $I3(\mathcal{P})$, suggesting that in general perfect states are local maxima of $I3(\mathcal{P})$ for these partitionings. Furthermore we explored the effect of different measurements on one and two of the qubits of the system, but even in this case we did not get any new state with positive value of $I3(\mathcal{P})$. It would be interesting to explore these results for larger systems, higher dimensional generalizations of the constituents and different classes of operations.

Finally, considering also the results from investigations of GHZ$_N$ states, it seems natural to expect that some amount of $4$-partite quantum entanglement is really crucial for the violation of monogamy of mutual information. Unfortunately, no measure of $4$-partite quantum entanglement for mixed state is available to investigate this expectation quantitatively.

\paragraph{The tripartite information as a parameter for scrambling}

Since perfect states might be thought as the result of scrambling, and they correspond to global minima of $I3(\mathcal{P})$, it was proposed in \citep{Hosur:2015ylk} that the tripartite information can be used as a parameter for scrambling. In our analysis of perfect states, we showed that for some permutation of the constituents of the system, the same value of $I3(\mathcal{P})$ can in principle be attained by products of perfect states of smaller systems. Since these product states are not perfect states of the larger system, one can conclude that the value of $I3(\mathcal{P})$ can be an appropriate measure of scrambling only under the assumption that the state under consideration is completely symmetric under permutations of the qubits. We propose that in general, as a measure of scrambling, one should use instead the average of the tripartite information ($\overline{I3(\mathcal{P})}$) over all possible permutations of the qubits.

Furthermore, since perfect states do not always exist, one can ask if for a given value of $N$, the state which contain the maximal possible amount of entanglement for all bipartitions (MMES) is the minimizer of $I3(\mathcal{P})$.  A counterexample to this expectation seems to derive from the highly entangled state of $8$ qubits found in \citep{Zha:2012aa}, which is conjectured to be a MMES state. We showed that the value of $I3(\mathcal{P})$ obtained for this state is smaller than the one obtained from the product of two copies of MMES of $4$ qubits. On the contrary, when we take the average of $I3(\mathcal{P})$ over all permutations of the qubits, the situation is reversed. This is a further argument in support of our proposal that $\overline{I3(\mathcal{P})}$ is a more appropriate parameter for scrambling.

\acknowledgments 

It is a great pleasure to thank Mukund Rangamani for many useful discussions and for comments on a preliminary draft. I also thank the hospitality of SITP at Stanford University where this project was initiated and in particular the Center for Quantum Mathematics and Physics (QMAP) at UC Davis, where most of this work was realized. Finally, I thank the APC at University of Paris 7, DAMTP at Cambridge University, NORDITA in Stockholm and the Niels Bohr Institute in Copenhagen, for hospitality during the last stages of this project. The research visit at Stanford and UC Davis was supported by the FQXi grant ``Measures of Holographic Information'' (FQXi-RFP3-1334).

\bibliographystyle{JHEP}
\bibliography{I3}

\end{document}